\documentclass[prd,preprint,nofootinbib,showpacs]{revtex4-1}
\pdfoutput=1
\usepackage{graphicx}
\usepackage{hyperref}
\usepackage{amsfonts}
\usepackage{amsmath,amssymb}
\usepackage{bm}
\usepackage{color}

\def\be{\begin{equation}}
\def\ee{\end{equation}}
\def\bea{\begin{eqnarray}}
\def\eea{\end{eqnarray}}
\def\ba{\begin{array}}
\def\ea{\end{array}}

\begin{document}

\title{Strongly Coupled  Striped Superconductor with Large Modulation}

\author{Jimmy A. Hutasoit}
\email{jimmy.hutasoit@mail.wvu.edu} 
\affiliation{Department of Physics, West Virginia University, Morgantown, West Virginia 26506}

\author{Suman Ganguli}
\email{sgangul3@tennessee.edu}
\author{George Siopsis}
\email{siopsis@tennessee.edu}
\author{Jason Therrien}
\email{jtherrie@tennessee.edu}
\affiliation{Department of Physics and Astronomy, The University of Tennessee, Knoxville, Tennessee 37996-1200}

\date{\today} 

\begin{abstract}
Using gauge/gravity duality, we analytically calculate properties of a strongly coupled striped superconductor, with the charge density wave sourced by a modulated chemical potential, in the large modulation wavenumber $Q$ limit. In the absence of a homogeneous term in the chemical potential, we show that the critical temperature scales as a negative power of $Q$ for scaling dimensions $\Delta < \frac{3}{2}$, whereas for $\Delta > \frac{3}{2}$, there is no phase transition above a certain critical value of $Q$. The condensate is found to scale as a positive power of $Q$ such that the gap is proportional to $Q$. We discuss how these results change if a homogeneous term is added to the chemical potential.
We compare our analytic results with numerical calculations whenever the latter are available and find good agreement.
\end{abstract}

\maketitle


\section{Introduction}
High temperature superconductors, or high $T_c$ superconductors, are materials whose superconducting transition temperature is higher than 30 $K$.  Even though such materials were discovered in the late 1980s, a sufficiently accurate theoretical understanding of such materials has been lacking (for a review with a strong emphasis on the theoretical issues, see for example, Ref. \cite{Carlson:fk}). The difficulty arises from the fact that these high temperature superconductors are strongly coupled \cite{Dagotto:uq}. With the lack of a general theoretical framework in quantum many-particle physics that deals with fermions at finite density, we face the so-called fermion sign problem, in which we cannot rely on brute force techniques using lattice models to help us solve this problem at strong coupling.  However, recently, by borrowing an idea that comes from string theory, i.e., gauge/gravity duality, progress has been made in our understanding of these strongly coupled superconductors. 

In gauge/gravity duality, the strongly coupled condensed matter systems are mapped to a weakly coupled gravitational theory on a spacetime with negative cosmological constant, or the so-called anti de-Sitter (AdS) spacetimes. In particular, the study of high $T_c$ superconductor is mapped to a study of Einstein-Maxwell-scalar theory living in four-dimensional AdS black holes and the superconducting phase transition is understood as the (unstable) black holes forming scalar 'hairs.' For a review on applications of gauge/gravity duality in condensed matter physics, see Ref. \onlinecite{Sachdev:2011fk} and references therein. For analytic studies of homogeneous holographic superconductor, see for example, Refs. \cite{Siopsis:2010uq,Siopsis:2010fk}.

Further complication that arises in the high temperature superconductors, such as cuprates and iron pnictides, is the competing orders that are related to the breaking of the lattice symmetries. At first, these orders seem to be unrelated to superconductivity, however, a study of the effect of inhomogeneity of the pairing interaction in a weakly coupled BCS system \cite{Martin:2005fk} as well as numerical studies of Hubbard models \cite{Hellberg:uq,White:kx} suggest that inhomogeneity might play a role in high $T_c$ superconductivity. Furthermore, the recent discovery of transport anomalies in ${\rm La}_{2-x} {\rm Ba}_x {\rm CuO}_4$, which are particularly prominent for $x = 1/8$ \cite{Li:2007}, might be explained under the assumption that this cuprate is a ``striped" superconductor \cite{Berg:2009fk}. Other studies using mean-field theory have also shown that unlike the homogeneous superconductor, the striped superconductor exhibits the existence of a Fermi surface in the ordered phase \cite{Baruch:2008uq,Radzihovsky:2008kx} and its complex sensitivity to quenched disorder \cite{Berg:2009fk}.

Therefore, it is of great value to have comprehensive models that incorporate not only strong coupling, but also inhomogeneity. The goal will be to have the competing orders emerging dynamically, however, in this article, we will follow Ref. \onlinecite{Flauger:2011vn} where the inhomogeneity is introduced via a modulated chemical potential, with wavenumber $Q$. Further work on the physical properties of our model is needed to determine its applicability to striped superconductors realized in Nature.

We are particularly interested in studying the behavior of the system at large $Q$. For non-vanishing homogeneous part of the chemical potential, the behavior of the leading terms of the observables are identical to those of a homogeneous superconductor with some small $Q$-dependent corrections. In this paper, we calculate these subleading terms and we find that they are not exponentially suppressed as previously suggested, but instead exhibit a power law behavior. When the homogeneous part of the chemical potential vanishes, we find a very different qualitative behavior compared to the homogeneous case. In particular, the critical temperature vanishes when the anomalous dimension of the condensate is larger than $3/2$. For values of the dimension below $3/2$, we show that the critical temperature scales as a negative power of $Q$ whereas the condensate scales as a positive power of $Q$ such that the gap is proportional to $Q$. This indicates that, perhaps, the underlying mechanism is not based on a correlation length as with BCS superconductors.\footnote{We thank Stefanos Papanikolaou for a discussion on this point.} It would be interesting to further study the properties of these superconductors, including two and higher-point correlation functions to better understand the role of the correlation length. 

Our discussion is organized as follows. In Section \ref{sec:2}, we set up the equations that govern the system. In Section \ref{sec:3}, we calculate the critical temperature in the large $Q$ limit. In Section \ref{sec:4}, we discuss the behavior of the condensate including subleading terms. Finally, in Section \ref{sec:5}, we summarize our conclusions.

\section{Set-up}
\label{sec:2}
The minimal requirement to study strongly coupled superconductor using gauge/gravity duality is to have a scalar field and a $U(1)$ gauge field living in a spacetime with negative cosmological constant. The scalar field is dual to a scalar order parameter of the superconductor, i.e., the condensate, while the $U(1)$ gauge field is dual to the current in the condensed matter system. 

To study the strong coupling regime of the superconductor, we only need to study the gravity theory at the classical level. In particular, we are interested in finding solutions to the classical equations of motion whose boundary values are related to the parameters of the superconductor.

The equation of motion for the scalar field is
\be\label{eq:eompsi}
- \frac{1}{\sqrt{-g}}D_a\left(\sqrt{-g}g^{ab}D_b\Psi\right) + \frac{1}{2} \frac{\Psi}{|\Psi|} V'(|\Psi|) = 0 \,,
\ee
while the equation for the U(1) gauge field is
\be\label{eq:eomA}
\frac{1}{\sqrt{-g}}\partial_a\left(\sqrt{-g} F^{ab}\right) = i g^{ba} \left[\Psi^* D_a\Psi - \Psi (D_a\Psi)^* \right]
\,,
\ee
and Einstein's equations are
\begin{multline}\label{eq:einst}
R_{ab} - \frac12g_{a b} R - \frac{3}{L^2}g_{ab} =8\pi G_N\Big[ F_{a c} F_b{}^c - \frac{1}{4}g_{a b} F^{cd} F_{cd} \\
 - g_{a b}
D_a\Psi(D_b\Psi)^* + \left[D_a\Psi (D_b\Psi)^* + a \leftrightarrow b \right]- g_{a b} V(|\Psi|)\vphantom{\frac14}\Big] \,.
\end{multline}
Here, $D_a = \partial_a - iqA_a$, where $a$ is the space-time index.

In the following, we will consider the potential to be quadratic, with the coefficient $m$ to be the mass of the scalar field. Furthermore, we will neglect the backreaction of the scalar field and the gauge field onto the metric. This is valid when scale invariant quantities, such as $G_N\mu^2$, where $\mu$ is the chemical potential, are small. As usual, this can be achieved by sending $G_N$ to zero while keeping everything else fixed.

Therefore, the problem of studying a strongly coupled superconductor with an effective two spatial dimensions is reduced to studying the dynamics of a scalar field with mass $m$, coupled to a $U(1)$ gauge field in the background of a 3+1-dimensional AdS Schwarzschild black hole with the following metric
\be
ds^2 = - r^2 \,h(r)\, dt^2 + \frac{dr^2}{r^2 \,h(r)} + r^2 \,d\vec{x}^2,
\ee
with
\be
h = 1 - \frac{r_+^3}{r^3},
\ee
in units in which the AdS radius is unity ($L=1$). The radius of the horizon is $r_+$ and the Hawking temperature, which is equivalent to the temperature of the strongly coupled system, is given by
\be
T = \frac{3}{4 \pi} r_+. \label{deftemp}
\ee
In this coordinate system, the boundary is at infinity.
 
By fixing the gauge, we can set $A_x = A_y = A_r = 0$ with non-zero electrostatic scalar potential $A_t (r) \equiv A(r)$. The equation of motion is then given by
\bea
A''-\frac{2\Psi^2}{z^2 h} A + \frac{\partial_x^2 A}{r_+^2 h} = 0, \label{Aeom}\\
\Psi'' + \left(\frac{h'}{h} - \frac 2z \right) \Psi' + \left( \frac{A^2}{r_+^2h^2} - \frac{m^2}{z^2 h} \right) \Psi + \frac{\partial_x^2  \Psi}{r_+^2 h} = 0,
\eea
where we have set $q=1$ and performed a coordinate transformation
\be
z = \frac{r_+}{r},
\ee
under which $h(r)$ becomes $h = 1 - z^3$, the boundary is at $z=0$, while the horizon is at $z=1$. Here $'$ denotes a derivative with respect to $z$.

The boundary behaviors of the fields are related to the observables in the strongly coupled theory as follows
\bea
A(z\rightarrow 0) &=& \mu - \frac{\rho}{r_+} \, z,  \\
\Psi(z\rightarrow 0) &=& \frac{\langle {\cal O}_{\Delta} \rangle}{\sqrt{2}} \, \frac{z^{\Delta}}{r_+^{\Delta}},
\eea
where $\mu$ is the chemical potential, $\rho$ is the charge density and $\langle {\cal O}_{\Delta} \rangle$ is the expectation value of the condensate of the strongly coupled system. As stated before, $r_+$ is related to the temperature via Eq.\ (\ref{deftemp}). 

To study the effect of inhomogeneity, we are going to consider the following electrostatic potential 
\be
A(z,x) = \mu  \sum_{n \geq 0} \delta^{(n)} \, A^{(n)}(z) \, \cos nQx,
\ee
with
\be
\sum_{n \geq 0} \delta^{(n)} = 1,
\ee
subject to the boundary conditions $A^{(n)}(0) = 1$ and $A^{(n)} (1) = 0$. 

This set-up is equivalent to the charge density wave (CDW) being sourced by a modulated chemical potential. We are particularly interested in the regime where $Q$ is larger compared to the temperature scale. 

\section{At the Critical Temperature}
\label{sec:3}
At the critical temperature $T_c$ and above, the order parameter vanishes, and thus $\Psi =0$. Therefore,
\be
{A^{(n)}}'' - \frac{n^2\, Q^2 \,A^{(n)}}{r_+^2 h} = 0.
\ee
For $n = 0$, the solution is
\be
A^{(0)} =  1-z,
\ee
while for $n>0$, we can solve the equation by first rescaling the radial coordinate $\tilde{z} = \frac{nQ}{r_+} z$ and then expanding the equation of motion formally in $\frac{r_+}{nQ}$. We obtain
\be\label{eqexpansion}
\partial_{\tilde{z}}^2 A^{(n)} = \left(1 + \frac{r_+^3}{n^3 Q^3} \tilde{z}^3 + \cdots \right) {A^{(n)}},
\ee
which is an expansion around the boundary $z=0$. This expansion is only valid for $z \lesssim \frac{r_+}{nQ}$, which covers a small part of the entire range of $z$ if $Q$ is large compared to the temperature. However, for $z \gtrsim \frac{r_+}{nQ}$, the potential is exponentially small, so the only physically significant range of $z$ is the one in which the formal expansion \eqref{eqexpansion} is valid.

Indeed, the leading-order solution is 
\be\label{eqlead}
A^{(n)}  = \cosh \tilde{z} - \coth \frac{nQ}{r_+} \, \sinh \tilde{z} = \frac{\sinh \frac{nQ}{r_+} (1-z)}{\sinh \frac{nQ}{r_+}}~.
\ee
The first-order corrections are of order $(nQ/r_+)^{-3}$ and therefore negligible at large $Q$ for $z \lesssim \frac{r_+}{nQ}$, whereas for $z \gtrsim \frac{r_+}{nQ}$, the potential \eqref{eqlead} clearly vanishes exponentially. As noted in Ref. \onlinecite{Flauger:2011vn}, this analytical expression agrees well with the numerical solution even for low $Q$. In fact, as $Q\to 0$, \eqref{eqlead} becomes exact ($A^{(n)} \to 1-z$).

In the following, we shall concentrate on the case with only two Fourier modes. In other words, $\delta^{(n)} =0$ for $n>1$, and therefore, let us set $\delta^{(1)} = \delta$ and $\delta^{(0)} = 1 - \delta$.

Let us define
\be
\Psi = \frac{\langle {\cal O}_{\Delta} \rangle^{(0)}}{\sqrt{2}} \, \frac{z^{\Delta}}{r_+^{\Delta}} \, \sum_{n  \geq 0} F^{(n)}(z) \, \cos nQx, \label{defmode}
\ee
with $F^{(0)}(z=0)=1$, such that the expectation value of the condensate is
\be
\langle {\cal O}_{\Delta} \rangle = \sum _{n  \geq 0} \,  \langle {\cal O}_{\Delta} \rangle^{(n)} \cos n Q x,
\ee
with 
\be
\langle {\cal O}_{\Delta} \rangle^{(n)} = \langle {\cal O}_{\Delta} \rangle^{(0)} \, F^{(n)}(z=0).
\ee 
Here, $\Delta = \Delta_\pm = \frac{3}{2} \pm \sqrt{\frac{9}{4} + m^2}$ is the dimension of the condensate operator ${\cal O}_{\Delta}$. We shall examine the range
\be
\frac12 < \Delta < 3,
\ee
where $\Delta > \frac{3}{2}$ $\left(\Delta < \frac{3}{2}\right)$ for $\Delta = \Delta_+$ ($\Delta = \Delta_-$), corresponding to masses in the range $0 > m^2 \ge -\frac{9}{4}$.

\subsection{Variational Method}

\begin{figure}[htbp]
\begin{center}
\includegraphics[width= 10 cm, keepaspectratio=true]{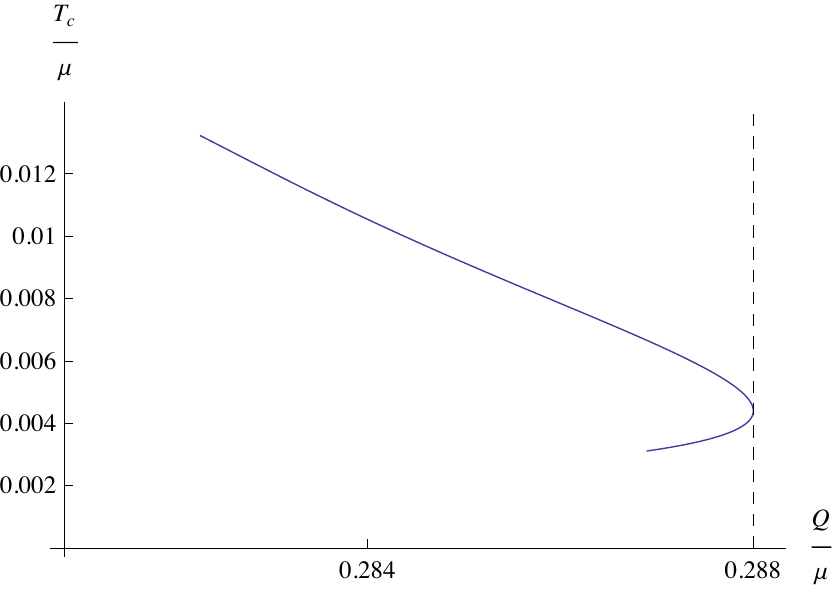}
\caption{Critical temperature $T_c$ for $\delta =1$ and $\Delta = 2$ obtained numerically. For $\Delta>3/2$, there is a critical modulation $Q^{\ast}$, at which $T_c$ discontinuously drops to zero. Here, $Q^{\ast} = 0.288\, \mu$. The lower branch in $Q < Q^{\ast}$ regime is unstable. }
\label{drop}
\end{center}
\end{figure}

The equation of motion for the scalar field can be written as
\bea
h z^{2\Delta - 2} \,{F^{(n)}}'' &+& (h z^{2\Delta - 2})' \,{F^{(n)}}' + \left[ (h-1) \Delta^2 - n^2 Q_c^2 z^2 \right]\,z^{2 \Delta -4}\, {F^{(n)}} \nonumber \\
&=& - \mu_c^2 \, \frac{z^{2 \Delta -2}}{h} \Bigg[(1-\delta)^2 \, {A^{(0)}}^2 F^{(n)} - \delta (1- \delta) A^{(0)} A^{(1)} \left(F^{(n-1)} + F^{(n+1)} \right)  \nonumber \\
& & \qquad \qquad  \qquad + \, \frac{\delta^2}{4} {A^{(1)}}^2 \left( F^{(n-2)} + 2 F^{(n)} + F^{(n+2)} \right)  \Bigg],
\eea
where $\mu_c$ ($Q_c$) is the chemical potential (wavenumber) in units of the horizon radius at the critical point,
\be
\mu_c = \frac{\mu}{r_{+c}} \ \ , \ \ \ \ Q_c = \frac{Q}{r_{+c}} ~.
\ee
Looking at the $n=0$ equation, the eigenvalue $\mu_c$ minimizes the expression
\be
\mu_c^2 = \frac{\int_0^1 dz \, z^{2\Delta - 4}\, \{h z^2 ({F^{(0)}}')^2 -  (h-1) \Delta^2 \, (F^{(0)})^2 \}}{\int_0^1 dz \,\frac{z^{2 \Delta -2}}{h} F^{(0)} \{(1-\delta)^2  {A^{(0)}}^2 F^{(0)} - 2 \delta  (1- \delta)  A^{(0)} A^{(1)} F^{(1)} +  \frac{\delta^2}{2} {A^{(1)}}^2 \left(F^{(0)} + F^{(2)} \right)  \}}.
\ee
Since at large $Q$, it is expected (and numerical study confirms) that  $F^{(0)} \gg F^{(1)}, F^{(2)}$, the above expression reduces to
\bea\label{eqlambda}
\mu_c^2 &=& \frac{\int_0^1 dz \, z^{2\Delta - 4}\, \{h z^2 ({F^{(0)}}')^2 -  (h-1) \Delta^2 \, (F^{(0)})^2 \}}{\int_0^1 dz \,\frac{z^{2 \Delta -2}}{h} (F^{(0)})^2 \{(1-\delta)^2  {A^{(0)}}^2  +  \frac{\delta^2}{2} {A^{(1)}}^2 \}}  \nonumber \\
&\approx&  \frac{\int_0^1 dz \, z^{2\Delta - 4}\, \{h z^2 ({F^{(0)}}')^2 - (h-1) \Delta^2 \, (F^{(0)})^2 \}}{\int_0^1 dz \,\frac{z^{2 \Delta -2}}{h} \,(1-\delta)^2 \, {A^{(0)}}^2 (F^{(0)})^2   +  \frac{\Gamma(2\Delta-1)}{2^{2\Delta}}\, \frac{\delta^2}{Q_c^{2 \Delta -1}} }  \eea
In the absence of a homogeneous term in the chemical potential ($\delta = 1$), Eq.\ (\ref{eqlambda}) becomes
\be
\mu_c^2 =   \frac{2^{2\Delta}}{\Gamma(2\Delta-1)}\, Q_c^{2 \Delta -1} 
\, \int_0^1 dz \, z^{2\Delta - 4}\, \{h z^2 ({F^{(0)}}')^2 - (h-1) \Delta^2 \, (F^{(0)})^2 \} \ee
Using the trial function $F^{(0)} = 1- \alpha z^2$, we obtain
\be \mu_c^2 = \frac{2^{2\Delta}}{\Gamma(2\Delta-1)}\, \left(\frac{2 \Delta^2 - 3 \Delta +6}{2 (2 \Delta +1)} \, \alpha^2 - \frac{\Delta^2}{\Delta + 1} \, \alpha + \frac{\Delta}{2} \right) Q_c^{2 \Delta -1}~. 
\ee
The minimum is attained at
\be
\alpha = \frac{\Delta^2 (2 \Delta +1)}{(\Delta+1) \, (2 \Delta^2 - 3 \Delta + 6)},
\ee
which results in the estimate
\be
\mu_c^2 = \frac{2^{2\Delta-1}\Delta}{\Gamma(2\Delta-1)}\, \frac{2 \Delta^2+9 \Delta +6}{(\Delta+1)^2 (2 \Delta^2 - 3 \Delta + 6)} \, Q_c^{2\Delta -1}. \label{varspecialmu}
\ee
Recall that for the above derivation to be valid, we needed $Q_c \gtrsim 1$ (i.e., $Q \gtrsim T_c$). Therefore, $\mu_c \gtrsim 1$. Since
\be \frac{Q}{\mu} = \frac{Q_c}{\mu_c} \sim \mu_c^{\frac{3-2\Delta}{2\Delta -1}} \ee
it follows that for $\Delta > \frac{3}{2}$, $Q/\mu \lesssim 1$, suggesting that there is a critical value of $Q$ above which there is no phase transition. This sudden drop in the critical temperature has been observed in the numerical studies, see Fig. \ref{drop}.

For $\Delta < \frac{3}{2}$, we deduce the estimate of the critical temperature,
\be \frac{T_c}{\mu} = \frac{3}{4\pi}
\left[ \frac{2^{2\Delta-1}\Delta}{\Gamma(2\Delta-1)}\, \frac{2 \Delta^2+9 \Delta +6}{(\Delta+1)^2 (2 \Delta^2 - 3 \Delta + 6)} \right]^{\frac{1}{2\Delta-3}} \left( \frac{Q}{\mu} \right)^{-\frac{2\Delta -1}{3-2\Delta }}~. \label{varspecial}
\ee
For $\Delta =1$, we have $\mu_c^2 = 1.7\, Q_c$ and
\be
\frac{T_c}{\mu} =  \frac{0.14}{Q/\mu}. \label{vartc}
\ee
Thus the critical temperature has a power law behavior in the large $Q$ limit. Of course, this is only an estimate of $T_c$. Nevertheless, the exponent of $Q$ is confirmed by the perturbative method to be discussed later, which yields not only the exponent but also the multiplicative coefficient accurately in the case $\delta =1$. We shall also compare this with the numerical results to assess the accuracy of the results of the variational method.

In the presence of a homogeneous term in the chemical potential ($\delta \ne 1$), Eq.\ (\ref{eqlambda}) becomes
\bea
\mu_c^2 &=&  \frac{\int_0^1 dz \, z^{2\Delta - 4}\, \{h z^2 ({F^{(0)}}')^2 -  (h-1) \Delta^2 \, (F^{(0)})^2 \}}{\int_0^1 dz \,\frac{z^{2 \Delta -2}}{h} \,(1-\delta)^2 \, (1 - z)^2 \, (F^{(0)})^2} \nonumber \\
& & \, -  \,  \frac{\int_0^1 dz \, z^{2\Delta - 4}\, \{h z^2 ({F^{(0)}}')^2 -   (h-1) \Delta^2\, (F^{(0)})^2 \}}{\left[\int_0^1 dz \,\frac{z^{2 \Delta -2}}{h} \,(1-\delta)^2 \, (1 - z)^2 \, (F^{(0)})^2 \right]^2} \, \frac{\Gamma(2\Delta-1)}{2^{2\Delta}}\, \frac{\delta^2}{Q_c^{2 \Delta -1}} , \label{lambda}
\eea
where we have approximated the second integral in the denominator by evaluating the integrand near the horizon. We note that the subleading term is not exponentially suppressed as suggested, based on numerical calculations \cite{Flauger:2011vn}, but is a power law in $Q$.
This is to be contrasted with the $1/\log Q$ behavior seen in the weak coupling BCS calculation \cite{Martin:2005fk}. 

Since the second term in the last line is subleading, we can find the eigenvalue $\mu_c$ by first minimizing the leading term using a trial function
\be
F^{(0)} = 1 - \alpha z^2.
\ee
For $\Delta = 1$, the minimum is attained at $\alpha \approx 0.24$, which yields
\be
\mu_c^2 = \frac{1.27}{\left(1-\delta\right)^2} -  \frac{0.94\,\delta^2}{(1-\delta)^4 \, Q_c}. \label{vargeneral}
\ee
The critical temperature is 
\be
\frac{T_c}{\mu} \approx 0.21 \, (1-\delta) +  \frac{0.078  \,\delta^2}{Q/\mu}.
\ee

We note that in obtaining the last line of Eq.\ (\ref{lambda}), we have assumed that the first term of the denominator is much greater than the second one. Obviously, this assumption is not always valid. In particular, it is easy to see that the limit $\delta \rightarrow 1$ does not commute with the large $Q$ limit. 

\subsection{Perturbative Method}

To gain more insight, let us solve the equation of motion for $F^{(0)}$ by treating the electrostatic potential as a perturbation to a leading order solution, which is nothing but that of a scalar field on an AdS Schwarzschild black hole without any background $U(1)$ gauge field
\be
{F_0^{(0)}}'' + \left(\frac{h'}{h} + \frac{2 (\Delta - 1)}{z}\right) {F_0^{(0)}}' - \frac{\Delta^2 z}{h} {F_0^{(0)}} = 0.
\ee
Here, ${F_0^{(0)}}$ is the leading term of the solution. Strictly speaking, this perturbation is valid only when $\delta \approx 1$, however, as we shall see in a bit, the result is a good approximation even when $\delta$ is far away from unity. 

The solution that satisfies the correct boundary conditions at $z = 0$ is
\be
{F_0^{(0)}} = \,_2F_1 \left(\frac{\Delta}{3}, \frac{\Delta}{3}; \frac{2 \Delta}{3}; z^3 \right),
\ee
where $_2F_1$ is the Gauss hypergeometric function. There is another solution, which corresponds to the solution with the correct boundary conditions for $\Delta \rightarrow  3 - \Delta$
\be
{\tilde{F}_0^{(0)}} = z^{3 - 2 \Delta} \, \,_2F_1 \left(\frac{3 -\Delta}{3}, \frac{3 - \Delta}{3}; \frac{2 (3 -\Delta)}{3}; z^3 \right).
\ee
Using perturbation theory, we obtain the next leading order solution, which is given by
\bea
F_1^{(0)} (z) &=& \frac{\mu_c^2}{3 - 2 \Delta} \left(  F_0^{(0)}(z) \int^z_0 \frac{dz'}{{z'}^{2-2 \Delta}} \tilde{F}_0^{(0)} \, \frac{{\cal A}}{h} \, {F}_0^{(0)} -   \tilde{F}_0^{(0)}(z) \int^z_0 \frac{dz'}{{z'}^{2-2 \Delta}} F_0^{(0)} \, \frac{{\cal A}}{h} \, {F}_0^{(0)}\right) \nonumber \\
&\equiv& \frac{\mu_c^2}{3 - 2 \Delta} \left(  F_0^{(0)}(z) \, \tilde{a}(z)  -   \tilde{F}_0^{(0)}(z) \, a(z) \right), \label{adef}
\eea
where
\be
{\cal A} (z) = (1-\delta)^2 \, {A^{(0)}}^2 + \frac{{\delta^2 \, A^{(1)}}^2}{2}~.
\label{adef1}
\ee

As both $F_0^{(0)}$ and $\tilde{F}_0^{(0)} $ diverge logarithmically at the horizon, we obtain the following singularity for the full solution
\bea
F^{(0)} (z \rightarrow 1) &=& F_0^{(0)} (z \rightarrow 1)  + F_1^{(0)} (z \rightarrow 1)  + \cdots\nonumber \\
&\approx& \log (1-z) \Bigg\{ \frac{\Gamma\left(\tfrac{2\Delta}{3}\right)}{\Gamma^2\left(\tfrac{\Delta}{3}\right)} \left[1 + \frac{\mu_c^2 \, \tilde{a}(1) }{3 - 2 \Delta} \right]  -  \frac{\Gamma\left(\tfrac{2(3-\Delta)}{3}\right)}{\Gamma^2\left(\tfrac{3-\Delta}{3}\right)} \frac{\mu_c^2 \, a (1)}{3 - 2 \Delta} \Bigg\},
\eea
where
\bea
a (1) &\approx& (1-\delta)^2 \, a^{(0)}_0 + \frac{\Gamma(2\Delta -1)}{2^{2\Delta}}\, \frac{\delta^2}{Q_c^{2\Delta -1}}, \nonumber \\
\tilde{a}(1)  &\approx& (1-\delta)^2 \, \tilde{a}^{(0)}_0 + \frac{1}{8} \, \frac{\delta^2}{Q_c^2}, \label{int1}
\eea
with
\bea 
a^{(0)}_0 &=& \int_0^1 \frac{dz}{{z}^{2-2 \Delta}} F_0^{(0)} \, \frac{(1-z)^2}{h} \, {F}_0^{(0)} , \nonumber \\
\tilde{a}^{(0)}_0 &=& \int_0^1 \frac{dz}{{z}^{2-2 \Delta}} \tilde{F}_0^{(0)} \, \frac{(1-z)^2}{h} \, {F}_0^{(0)}.  \label{a0}
\eea
In obtaining Eq.\ (\ref{int1}), we have approximated the subleading integrals by evaluating the integrand near the boundary $z = 0$. 

Requiring regularity at the horizon, we obtain
\bea
\frac{1}{\mu_c^2} &=& \frac{1}{3 - 2\Delta} \left( \frac{\Gamma^2\left(\tfrac{\Delta}{3}\right)}{\Gamma\left(\tfrac{2\Delta}{3}\right)} \, \frac{\Gamma\left(\tfrac{2(3-\Delta)}{3}\right)}{\Gamma^2\left(\tfrac{3-\Delta}{3}\right)} a(1) - \tilde{a}(1) \right) \nonumber \\
&=& \left(\frac{\Gamma^2\left(\tfrac{\Delta}{3}\right)}{\Gamma\left(\tfrac{2\Delta}{3}\right)} \, \frac{\Gamma\left(\tfrac{2(3-\Delta)}{3}\right)}{\Gamma^2\left(\tfrac{3-\Delta}{3}\right)} a^{(0)}_0 - \tilde{a}^{(0)}_0 \right) \frac{(1-\delta)^2}{3-2\Delta} +  \frac{\Gamma^2\left(\tfrac{\Delta}{3}\right)}{\Gamma\left(\tfrac{2\Delta}{3}\right)} \, \frac{\Gamma\left(\tfrac{2(3-\Delta)}{3}\right)}{\Gamma^2\left(\tfrac{3-\Delta}{3}\right)}\frac{\Gamma(2\Delta -1)}{2^{2\Delta}(3-2\Delta)}\frac{\delta^2}{Q_c^{2\Delta -1}}, \nonumber \\ \label{pertcomplete}
\eea
where we have dropped the term proportional to $1/Q_c^2$ on the last line. Both terms in Eq.\ (\ref{pertcomplete}) are small, but to connect this result with the result from variational method Eqs.\ (\ref{vargeneral}) and (\ref{varspecial}), we make two different approximations in which one is a lot larger than the other.

When the first term is significantly larger than the second one, for $\Delta =1$, we have
\be
\mu_c^2 = \frac{1.19}{(1-\delta)^2} - \frac{0.91 \, \delta^2}{(1-\delta)^4 \, Q_c},
\ee
which agrees well with Eq.\ (\ref{vargeneral}). We would like to emphasize that this agreement is valid even when $\delta$ is far away from unity.

When the second term is significantly larger than the first, we have
\be
\mu_c^2 = \frac{\Gamma\left(\tfrac{2\Delta}{3}\right)}{\Gamma^2\left(\tfrac{\Delta}{3}\right)} \, \frac{\Gamma^2\left(\tfrac{3-\Delta}{3}\right)}{\Gamma\left(\tfrac{2(3-\Delta)}{3}\right)} \frac{{2^{2\Delta}(3-2\Delta)}}{\Gamma(2\Delta -1)} \, \frac{Q_c^{2\Delta -1}}{\delta^2}. \label{pertlambda}
\ee
For $\Delta =1$, $\delta =1$, we then have $\mu_c^2 = 1.55 \,Q_c$ and
\be
\frac{T_c}{\mu} = \frac{0.15}{Q/\mu},
\ee
which agrees with the previous result obtained by variational method Eq.\ (\ref{vartc}).

We would like to note that for $\delta =1$ and $\Delta >3/2$, Eq. \ref{pertlambda} does not have any solutions. This is related to the fact that for $\delta =1$ and $\Delta >3/2$, there is a critical $Q^{\ast}$ such that $T_c(Q>Q^{\ast})=0$, as we have mentioned earlier.

Let us also comment on the $\Delta = 1/2$ unitarity limit, which is singular.
To approach it, we introduce a cutoff $\Lambda$ in $Q$-space. From Eq.\  (\ref{pertcomplete}), we have
\be
\frac{1}{\mu_c^2(Q)} - \frac{1}{\mu_c^2(\Lambda)} = \frac{\Gamma^2\left(\tfrac{\Delta}{3}\right)}{\Gamma\left(\tfrac{2\Delta}{3}\right)} \, \frac{\Gamma\left(\tfrac{2(3-\Delta)}{3}\right)}{\Gamma^2\left(\tfrac{3-\Delta}{3}\right)}\frac{\Gamma(2\Delta -1)}{2^{2\Delta}(3-2\Delta)}\,\delta^2 \left(Q^{1-2\Delta} - \Lambda^{1- 2\Delta}\right).
\ee
If $\Delta >1/2$, we can safely take the limit $\Lambda \rightarrow \infty$, in which $\Lambda^{1-2\Delta} \rightarrow 0$ and $\mu_c(\Lambda)$ has a finite limit. This is not so for the case of $\Delta=1/2$. Taking the limit $\Delta \rightarrow 1/2$, we have
\be
\lim_{\Delta \to \frac{1}{2}} \left[\frac{1}{\mu_c^2(Q)} - \frac{1}{\mu_c^2(\Lambda)} \right]=\frac{\Gamma^2\left(\tfrac{1}{6}\right)}{\Gamma\left(\tfrac{1}{3}\right)} \, \frac{\Gamma\left(\tfrac{5}{3}\right)}{\Gamma^2\left(\tfrac{5}{6}\right)}\frac{\delta^2}{4}\, \log \frac{\Lambda}{Q},
\ee
which shows that the limit $\Lambda \to \infty$ is not well defined and that for $\Delta =1/2$, the chemical potential is no longer a physical quantity. A well defined physical quantity would be
\be
\frac{d}{d\log Q}\left[\frac{1}{\mu_c^2(Q)} \right] = - \frac{\Gamma^2\left(\tfrac{1}{6}\right)}{\Gamma\left(\tfrac{1}{3}\right)} \, \frac{\Gamma\left(\tfrac{5}{3}\right)}{\Gamma^2\left(\tfrac{5}{6}\right)}\frac{\delta^2}{4}\, .
\ee

\begin{figure}[htbp]
\begin{center}
\includegraphics[width= 17 cm, keepaspectratio=true]{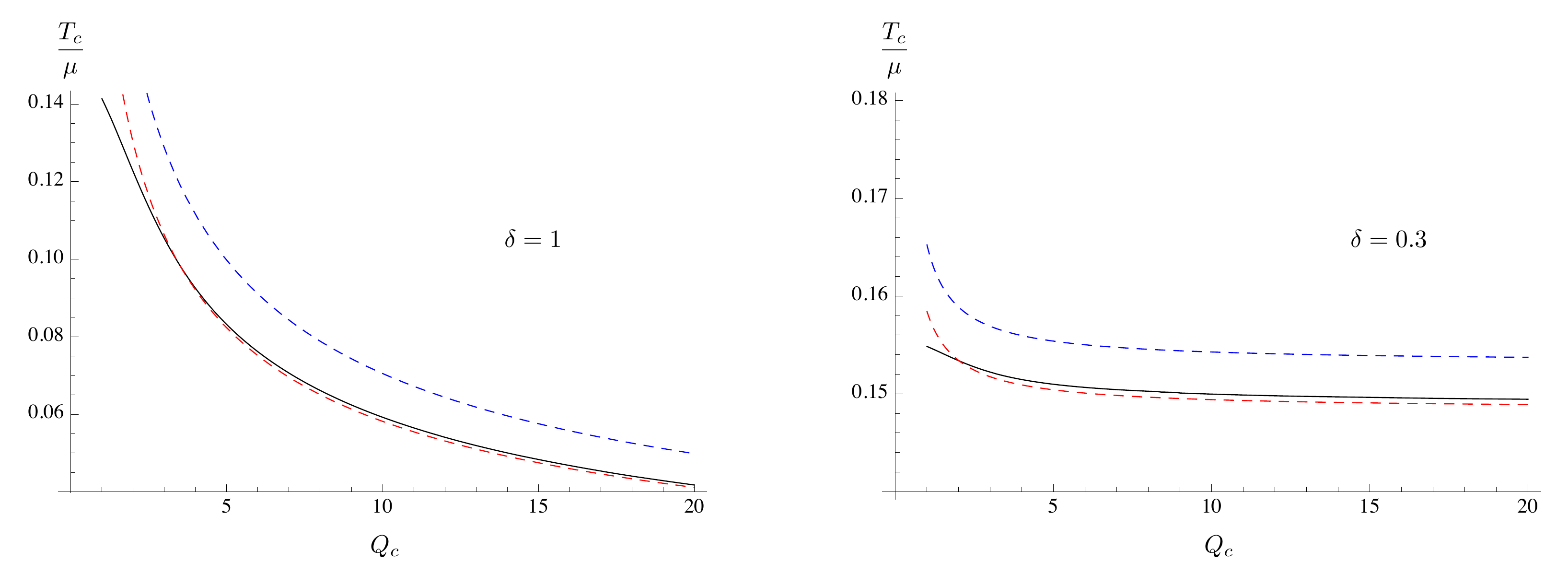}
\caption{The critical temperature as a function of $Q_c$ for $\Delta = 1$. The black lines are obtained by numerical analysis and the dashed red lines, which are closer to the numerical results, are obtained by perturbative method.}
\label{compare1}
\end{center}
\end{figure}

\begin{figure}[htbp]
\begin{center}
\includegraphics[width= 17 cm, keepaspectratio=true]{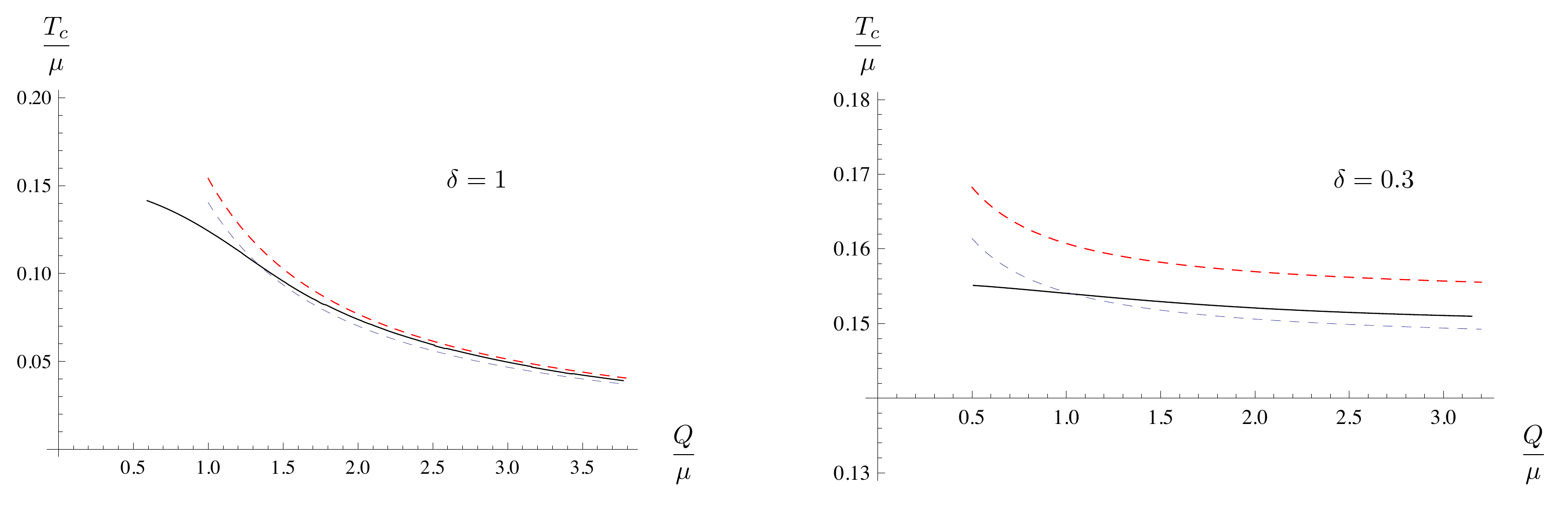}
\caption{The critical temperature as a function of $Q/\mu$ for $\Delta = 1$. The black solid lines, dashed red lines and the dashed blue lines are obtained by numerical analysis, perturbative method and variational method, respectively.}
\label{compare2}
\end{center}
\end{figure}

Lastly, let us compare the critical temperatures obtained by the variational and perturbative methods with the numerical result. This is depicted in Fig. \ref{compare1}. The agreement between the results obtained by analytical methods and the numerical results for $Q_c \gtrsim 3$ is remarkable. We also show the critical temperature as a function of $Q/\mu$ for different values of $\delta$ in Fig. \ref{compare2}.

\subsection{Higher Modes} \label{highmode}
\begin{figure}[htbp]
\begin{center}
\includegraphics[width=12 cm, keepaspectratio=true]{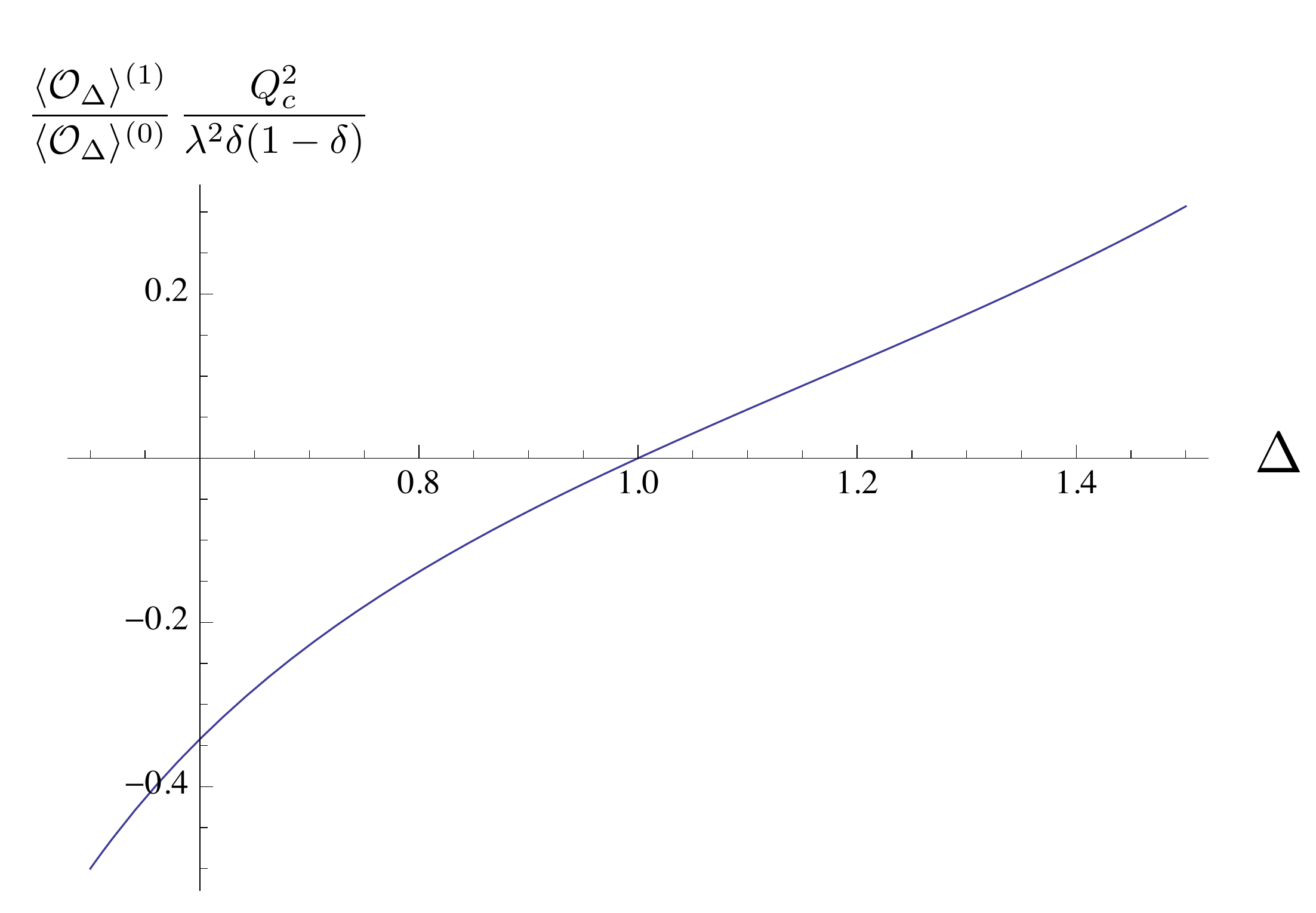}
\caption{The ratio $\frac{\langle {\cal O}_{\Delta}\rangle^{(1)}}{\langle {\cal O}_{\Delta}\rangle^{(0)}} \frac{Q_c^2}{\lambda^2 \delta (1-\delta)}$ as a function $\Delta$.}
\label{cond1}
\end{center}
\end{figure}

Using the perturbative method, we can also study the behavior of the subleading term of the condensate for $1/2<\Delta < 3/2$. For $\delta \ne 1$, the $n=1$ mode does not vanish and at sub-leading order, it is given by
\bea
F_1^{(1)} &=& F_0^{(1)} \left[\frac{\langle {\cal O}_{\Delta}\rangle^{(1)}}{\langle {\cal O}_{\Delta}\rangle^{(0)}} - \frac{\mu_c^2 \,\delta\,(1-\delta)}{3 - 2 \Delta} \int_0^z \frac{dz'}{{z'}^{2 - 2 \Delta}} \tilde{F}_0^{(1)} \frac{A^{(0)}A^{(1)}}{h}{F}_0^{(0)} \right]  \nonumber \\
& & \qquad \qquad \qquad +\,   \tilde{F}_0^{(1)} \, \frac{\mu_c^2 \, \delta \, (1-\delta)}{3 - 2 \Delta} \int^z_0 \frac{dz'}{{z'}^{2-2 \Delta}} F_0^{(1)} \frac{A^{(0)} A^{(1)}}{h}{F}_0^{(0)}, \label{pertcondsetupgeneral}
\eea
where $F_0^{(1)}$ and $\tilde{F}_0^{(1)}$ obey
\be
\partial_z \left(h\, z^{2\Delta -2}\, \partial_z F_0^{(1)}\right) - \left(z \Delta^2 + Q_c^2\right) z^{2\Delta -2}\, F_0^{(1)}= 0,
\ee
with the boundary conditions $F_0^{(1)} (z=0) =1$ and $\tilde{F}_0^{(1)} \approx z^{3- 2 \Delta}$. With $\mu_c^2$ given by Eq.\ (\ref{pertcomplete}), the boundary value 
\be
F^{(1)} (z=0) = \frac{\langle {\cal O}_{\Delta}\rangle^{(1)}}{\langle {\cal O}_{\Delta}\rangle^{(0)}}
\ee
is fully determined by demanding regularity at the horizon. To determine this, let us introduce the tortoise coordinate, defined as
\bea
r_{\ast} &=& (3-2\Delta) \int_0^z \frac{dz'}{h {z'}^{2\Delta -2}} \nonumber \\
&=& \frac{3-2 \Delta}{3} \, B\left(z^3; \frac{3-2 \Delta}{3}, 0\right),
\eea
where $B\left(x ; a, b\right)$ is the incomplete Euler beta function. We note that near the boundary $r_{\ast}(z\rightarrow0) = z^{3-2\Delta}$. In this tortoise coordinate, the equation of motion for $F_0^{(1)}$ and $\tilde{F}_0^{(1)}$ can be rewritten in the Schr$\ddot{\rm o}$dinger form
\be
- \partial_{r_{\ast}^2} F_0^{(1)} + V F_0^{(1)} = 0,
\ee
with the ``potential" $V$ given by
\bea
V &=& \frac{Q_c^2+ z \Delta^2}{\left(3-2\Delta\right)^2}\,h\,z^{4(\Delta-1)} \nonumber\\
&\approx&\left(\frac{Q_c}{3-2\Delta}\right)^2\,r_{\ast}^{\frac{4(\Delta-1)}{3-2\Delta}} \, \exp\left(- r_{\ast}^{\frac{3}{3-2\Delta}}\right).
\eea
Rescaling the tortoise coordinate $\tilde{r}_{\ast} = Q_c^{3 - 2 \Delta} \, r_{\ast}$, at large $Q_c$, we have
\be
- \partial_{\tilde{r}_{\ast}^2}  F_0^{(1)} +\left(\frac{1}{3-2\Delta}\right)^2\,\tilde{r}_{\ast}^{\frac{4(\Delta-1)}{3-2\Delta}} F_0^{(1)} = 0,
\ee
whose solutions are given in terms of modified Bessel functions
\be
 F_0^{(1)} = \frac{\pi}{\Gamma\left(\tfrac{3-2\Delta}{2}\right)} \,\frac{ 2^{\frac{2\Delta-3}{2}}}{\sin \tfrac{\pi (3-2\Delta)}{2}} \,\sqrt{\tilde{r}_{\ast}} \,\, I_{\tfrac{2\Delta-3}{2}}\left(\tilde{r}_{\ast}^{\frac{1}{3-2\Delta}}\right),
 \ee
and
\be
\tilde{F}_0^{(1)} = \frac{\Gamma\left(\tfrac{5-2\Delta}{2}\right)}{Q_c^{3-2\Delta}} 2^{\frac{3-2\Delta}{2}}\,\sqrt{\tilde{r}_{\ast}}\, \, I_{\tfrac{3-2\Delta}{2}}\left(\tilde{r}_{\ast}^{\frac{1}{3-2\Delta}}\right).
 \ee
These solutions satisfy the boundary conditions $F_0^{(1)} (r_{\ast}=0) =1$ and $\tilde{F}_0^{(1)} (r_{\ast}=0) = \tilde{r}_{\ast}$ but they diverge at the horizon. Their ratio at large $Q_c$ is then given by
\bea
\lim_{\tilde{r}_{\ast}\rightarrow \infty} \frac{\tilde{F}_0^{(1)}}{F_0^{(1)}} = \frac{\Gamma\left(\tfrac{5-2\Delta}{2}\right) \, \Gamma\left(\tfrac{3-2\Delta}{2}\right)}{\pi}\, 2^{3-2\Delta} \, \sin\left(\tfrac{\pi (3-2\Delta)}{2}\right)  \, Q_c^{2 \Delta -3}.
\eea
Substituting this into Eq.\ (\ref{pertcondsetupgeneral}) and demanding the full solution to the $n=1$ mode be regular at the horizon, we obtain
\be
\frac{\langle {\cal O}_{\Delta}\rangle^{(1)}}{\langle {\cal O}_{\Delta}\rangle^{(0)}} = \frac{\delta \,(1-\delta)}{3-2\Delta} \left[1- \frac{\Gamma\left(\tfrac{5-2\Delta}{2}\right) \, \Gamma\left(\tfrac{3-2\Delta}{2}\right)\,\Gamma\left(2\Delta-1\right)}{\pi \, 2^{2\Delta -3}}\, \sin\left(\tfrac{\pi (3-2\Delta)}{2}\right) \right]\frac{\mu^2}{Q^2} \left[ 1 + {\cal O}\left(\frac{1}{Q_c}\right) \right] ~. \label{pertcond1}
\ee
The leading order coefficient is plotted in Fig.\ \ref{cond1}. As this vanishes for $\Delta = 1$ and $\mu_c^2(\Delta =1)$ is finite, the leading order is ${\cal O}(\mu_c^2/Q_c^3)$.

\begin{figure}[htbp]
\begin{center}
\includegraphics[width=12 cm, keepaspectratio=true]{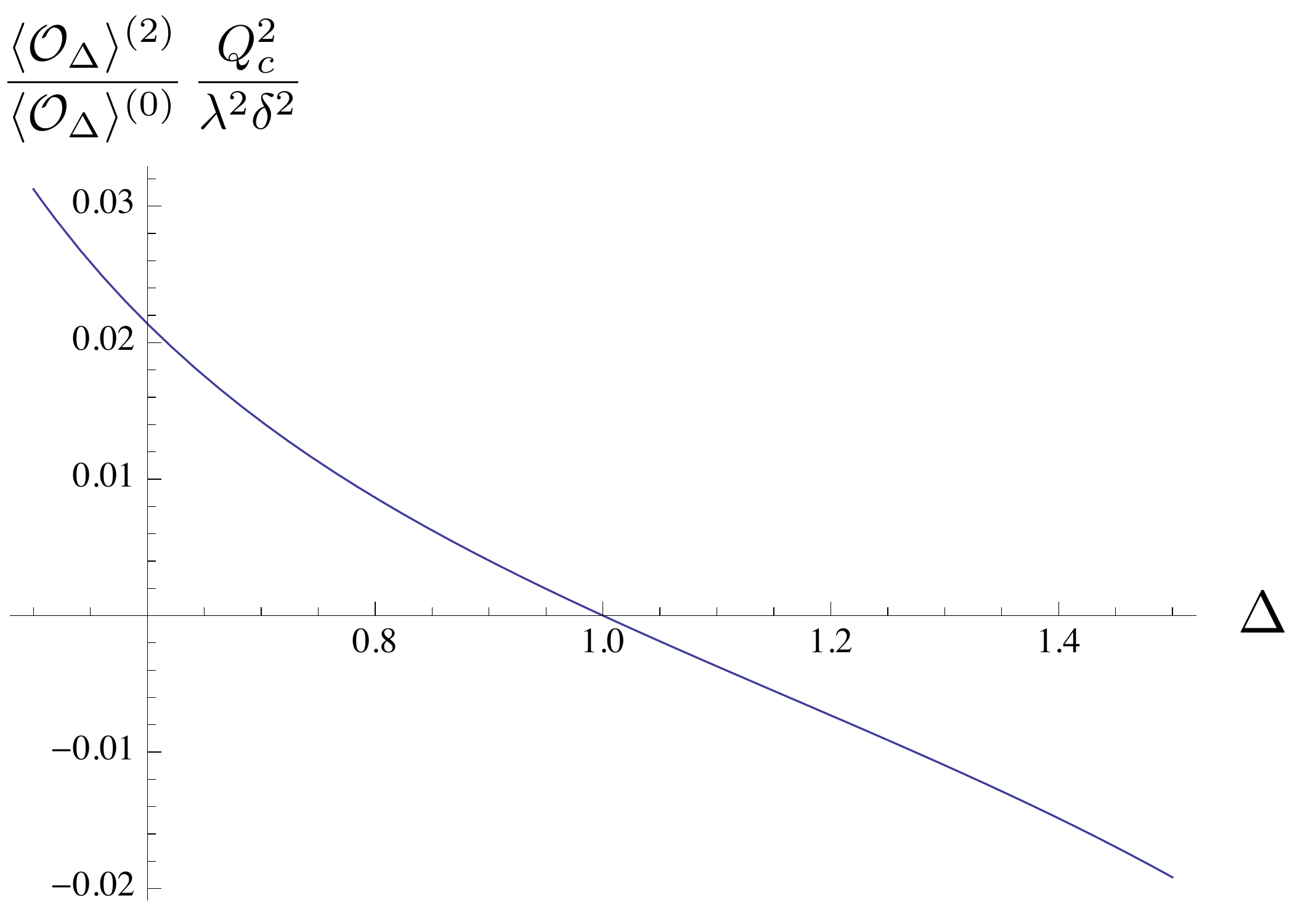}
\caption{The ratio $\frac{\langle {\cal O}_{\Delta}\rangle^{(2)}}{\langle {\cal O}_{\Delta}\rangle^{(0)}} \frac{Q_c^2}{\lambda^2 \delta^2}$ as a function $\Delta$.}
\label{NLOcond}
\end{center}
\end{figure}

Similarly, we  obtain
\be
\frac{\langle {\cal O}_{\Delta}\rangle^{(2)}}{\langle {\cal O}_{\Delta}\rangle^{(0)}} = \frac{\delta^2}{4 (3-2\Delta)} \left[\frac{\Gamma\left(\tfrac{5-2\Delta}{2}\right) \, \Gamma\left(\tfrac{3-2\Delta}{2}\right)\,\Gamma\left(2\Delta-1\right)}{\pi \, 2^{2\Delta -1}}\, \sin\left(\tfrac{\pi (3-2\Delta)}{2}\right) - \frac{1}{4}\right]\frac{\mu^2}{Q^2} \left[ 1+ {\cal O}\left(\frac{1}{Q_c}\right) \right]~. \label{pertcond2}
\ee
The dependence of this ratio on $\Delta$ is depicted in Fig.  \ref{NLOcond}. Again, we see that for $\Delta =1$, the bracketed quantity vanishes and the ratio is of order ${\cal O}(\mu_c^2/Q_c^3)$.

It is easy to see that for $\delta \ne 1$, when $\Delta \ne 1$
\be
\frac{\langle {\cal O}_{\Delta}\rangle^{(n)}}{\langle {\cal O}_{\Delta}\rangle^{(0)}} = {\cal O}\left(\frac{\mu^{2 \left[n/2\right]}}{Q^{2 \left[n/2\right]}}\right),
\ee
where $\left[n\right]$ denotes the largest integer $\leq n$, and
\be
\frac{\langle {\cal O}_{1}\rangle^{(n)}}{\langle {\cal O}_{1}\rangle^{(0)}} = {\cal O}\left(\frac{\mu^{2 \left[n/2\right]}\, T_c^{\left[n/2\right]}}{Q^{3 \left[n/2\right]}}\right),
\ee
for $\Delta =1$. For $\delta=1$, the odd modes vanish while the result for the even modes stays the same. This result confirms that at large $Q$, higher modes are negligible.

\section{Below the Critical Temperature}
\label{sec:4}

Below the critical temperature, $\Psi \ne 0$ and we have to include its contribution toward the electrostatic potential. From Eqs. \ref{Aeom} and \ref{defmode}, and by neglecting the higher modes, we have
\be
{A^{(n)}}'' - \left(\frac{n^2 \,Q^2}{r_+^2h}+\frac{{\langle {\cal O}_{\Delta} \rangle^{(0)}}^2 \, z^{2 (\Delta -1)}}{r_+^{2\Delta} h} {F^{(0)}}^2 \right)A^{(n)} = 0.
\ee
Rescaling $\tilde{z} = \frac{Q}{r_+}\,z$, we have
\be
\partial_{\tilde{z}}^2{A^{(n)}} - n^2 \,A^{(n)} = \frac{{\langle {\cal O}_{\Delta} \rangle^{(0)}}^2 \, \tilde{z}^{2 (\Delta -1)}}{Q^{2 \Delta}}  \, A^{(n)}, \label{backreact}
\ee
where we have dropped terms of order $1/Q^3$.

\subsection{$\Delta =1$}
The $\Delta =1$ case is a particularly easy one to solve and the solution is given by
\be
A^{(n)} = \frac{\sinh \frac{\sqrt{n^2 Q^2 + {\langle {\cal O}_{1} \rangle^{(0)}}^2}}{r_+} \,(1-z)}{\sinh \frac{\sqrt{n^2 Q^2 + {\langle {\cal O}_{1} \rangle^{(0)}}^2}}{r_+}}~. \label{alowtspec}
\ee
Substituting this into the equation of motion for the scalar field and demanding regularity at the horizon, we have
\be
\frac{r_+^2}{\mu^2} = \frac{\Gamma^3\left(\tfrac{1}{3}\right)}{3 \, \Gamma^3\left(\tfrac{2}{3}\right)} a(1) - \tilde{a}(1) ~, \label{rovermudef}
\ee
where $a(1)$ and $\tilde{a}(1)$ are defined in Eq.\ (\ref{adef}), but now with
$A^{(n)}$ ($n=0,1$) given by (\ref{alowtspec}).

For $\delta =1$, we obtain
\be
\frac{r_+}{\mu^2} = \frac{\Gamma^3\left(\tfrac{1}{3}\right)}{12 \, \Gamma^3\left(\tfrac{2}{3}\right)} \ \frac{1}{\sqrt{Q^2 + {\langle {\cal O}_{1} \rangle^{(0)}}^2}}~. \label{rovermuspecial1}
\ee
Multiplying by $\mu_c^2$, and solving for the condensate, we obtain
\be
\langle {\cal O}_{1} \rangle^{(0)} = Q \sqrt{\left( \frac{T_c}{T} \right)^2 -1}~, \label{rovermuspecial2}
\ee
showing that in the absence of a homogeneous term in the chemical potential, the condensate is proportional to the wavenumber $Q$.

Including a homogeneous term ($\delta \ne 1$) changes this behavior.
The case when the temperature is not too far away from the critical value, such that the expectation value of the condensate is small, will be treated when we study the case of general $\Delta$, but here let us assume that the condensate is large (low temperature regime). Then
\be
\frac{r_+}{\mu^2} = \frac{\Gamma^3\left(\tfrac{1}{3}\right)}{3 \, \Gamma^3\left(\tfrac{2}{3}\right)} \left[\frac{(1-\delta)^2}{2 \, \langle {\cal O}_{1} \rangle^{(0)}}  + \frac{\delta^2}{4\sqrt{Q^2 + {\langle {\cal O}_{1} \rangle^{(0)}}^2}}\right]. \label{rovermuspecial}
\ee
Multiplying by $\mu_c^2$, we obtain for $\delta \ne 1$ and $\langle \mathcal{O}_1 \rangle^{(0)} \lesssim Q$,
\be
\frac{T}{T_c^2} = 1.54 \left(\frac{1}{\langle {\cal O}_{1} \rangle^{(0)}} + \frac{\delta^2}{2(1-\delta)^2}\, \frac{1}{Q} + \dots \right) ~. \label{conddelta1}
\ee
Therefore,
\be
\langle {\cal O}_{1} \rangle^{(0)} = 6.4 \frac{T_c^2}{T} + \mathcal{O} (1/Q)~.
\ee
For $Q \lesssim \langle \mathcal{O}_1 \rangle^{(0)}$, we similarly obtain at leading order
\be \langle {\cal O}_{1} \rangle^{(0)} = 6.4 \frac{ 1 - 2\delta + \frac{3\delta^2}{2} }{(1-\delta)^2} \frac{T_c^2}{T}.
\ee

\subsection{Near $T_c$}
Let us now consider the case of general $\Delta$. For $n=0$, the solution that satisfies the boundary conditions is given in terms of the modified Bessel function of the second kind
\be
A^{(0)} = \frac{{\langle {\cal O}_{\Delta} \rangle^{(0)}}^{\frac{1}{2\Delta}}\,2^{\frac{2\Delta-1}{2\Delta}}}{\Gamma\left(\tfrac{1}{2\Delta}\right) \, r_+^{\frac{1}{2}} \, \Delta^{\frac{1}{2\Delta}}} \, \sqrt{z} \,\, K_{\frac{1}{2\Delta}}\left(\frac{\langle {\cal O}_{\Delta} \rangle^{(0)} \, z^{\Delta}}{\Delta\,r_+^{\Delta} } \right). \label{A0lowt}
\ee
For $n>0$, let us first consider a temperature not too far away below $T_c$, such that $\langle {\cal O}_{\Delta} \rangle^{(0)}$ is small and the right hand side of Eq.\ (\ref{backreact}) can be treated as perturbation. The solution is then expanded as
\be
A^{(n)} = A^{(n)}_0 + A^{(n)}_1 + \cdots,
\ee
where
\bea
A^{(n)}_0  &=& \frac{\sinh \frac{nQ}{r_+} (1-z)}{\sinh \frac{nQ}{r_+}}, \nonumber \\
A^{(n)}_1 &=& -\frac{{\langle {\cal O}_{\Delta} \rangle^{(0)}}^2 \, }{r_+^{2\Delta-1} \, n \,Q} \left( A^{(n)}_0 \int_0^1 dz' \,\frac{\sinh \frac{nQ}{r_+}z'}{(z')^{2 (1-\Delta)}}  \, A^{(n)}_0 - \int_z^1 dz' \,\frac{\sinh \frac{n Q}{r_+}(z'-z)}{(z')^{2 (1-\Delta)}} \, A^{(n)}_0  \right) \nonumber \\
&=& -  \frac{z^{2\Delta-1}}{2(2 \Delta-1)} \, \frac{e^{-nQz/r_+}}{nQ}\, \frac{{\langle {\cal O}_{\Delta} \rangle^{(0)}}^2}{r_+^{2\Delta-1}}. \label{gaugehighneartc}
\eea
For small $\langle {\cal O}_{\Delta} \rangle^{(0)}$, the zero mode of the electrostatic potential can also be expanded as
\be
A^{(0)} = 1-z - \frac{(2\Delta-1) z^{2\Delta+1} - (2\Delta+1) z^{2\Delta} +2z}{(2\Delta-1)\,2\Delta\,(2\Delta+1)} \, \frac{{\langle {\cal O}_{\Delta} \rangle^{(0)}}^2}{r_+^{2\Delta}}. \label{leadinggaugeneartc}
\ee
It is easy to see that by substituting $\Delta=1$ into the above equations, we recover the result of Eq.\ (\ref{alowtspec}) in this regime.

Substituting this into the equation of motion for the scalar and demanding regularity at the horizon, we obtain Eq.\ (\ref{rovermudef}), but now with
\be
{\cal A} = \frac{1}{2} \, \frac{\sinh^2 \frac{Q}{r_+}(1-z)}{\sinh^2 \frac{Q}{r_+}} -  \frac{z^{2\Delta-1}}{2(2 \Delta-1)} \, \frac{e^{-2Qz/r_+}}{Q}\, \frac{{\langle {\cal O}_{\Delta} \rangle^{(0)}}^2}{r_+^{2\Delta -1}},
\ee
for $\delta = 1$. We deduce
\bea
\frac{r_+^2}{\mu^2} &=&  \frac{\Gamma^2\left(\tfrac{\Delta}{3}\right)}{\Gamma\left(\tfrac{2\Delta}{3}\right)} \, \frac{\Gamma\left(\tfrac{2(3-\Delta)}{3}\right)}{\Gamma^2\left(\tfrac{3-\Delta}{3}\right)}\frac{1}{2^{2\Delta}(3-2\Delta)}\frac{r_+^{2\Delta -1}}{Q^{2\Delta -1}} \left[ \Gamma(2\Delta -1)- \frac{\Gamma(4\Delta -2)}{2\Delta -1} \frac{1}{2^{2 \Delta -1} } \, \frac{{\langle {\cal O}_{\Delta} \rangle^{(0)}}^2}{Q^{2\Delta}}\right]. \nonumber \\
\eea
Mulitplying by $\mu_c^2$, after some algebra, we obtain the condensate
\bea
\langle {\cal O}_{\Delta} \rangle^{(0)} &=& \sqrt{\frac{(2\Delta -1)2^{2\Delta-1}\Gamma(2\Delta-1)}{\Gamma(4\Delta -2)}}\,Q^{\Delta} \sqrt{1-\left( \frac{T}{T_c} \right)^{3-2\Delta}}.
\eea
showing that for $\delta =1$, in the large $Q$ limit,
\be \langle {\cal O}_\Delta \rangle^{1/\Delta} \propto Q. \ee
Similarly, in the case $\delta \ne 1$, we have
%
\bea
\frac{r_+^2}{\mu^2} &=& \left(\frac{\Gamma^2\left(\tfrac{\Delta}{3}\right)}{\Gamma\left(\tfrac{2\Delta}{3}\right)} \, \frac{\Gamma\left(\tfrac{2(3-\Delta)}{3}\right)}{\Gamma^2\left(\tfrac{3-\Delta}{3}\right)} a^{(0)}_0 - \tilde{a}^{(0)}_0 \right) \frac{(1-\delta)^2}{3-2\Delta} +  \frac{\Gamma^2\left(\tfrac{\Delta}{3}\right)}{\Gamma\left(\tfrac{2\Delta}{3}\right)} \, \frac{\Gamma\left(\tfrac{2(3-\Delta)}{3}\right)}{\Gamma^2\left(\tfrac{3-\Delta}{3}\right)}\frac{\Gamma(2\Delta -1)}{2^{2\Delta}(3-2\Delta)}\frac{\delta^2}{Q^{2\Delta -1}} \nonumber \\
&& \; + \left(\frac{\Gamma^2\left(\tfrac{\Delta}{3}\right)}{\Gamma\left(\tfrac{2\Delta}{3}\right)} \, \frac{\Gamma\left(\tfrac{2(3-\Delta)}{3}\right)}{\Gamma^2\left(\tfrac{3-\Delta}{3}\right)} a^{(0)}_1 - \tilde{a}^{(0)}_1 \right) \frac{(1-\delta)^2}{(2\Delta-3) \, \Delta\,(2\Delta-1)\,(2\Delta+1)} \, \frac{{\langle {\cal O}_{\Delta} \rangle^{(0)}}^2}{r_+^{2\Delta}},
\eea
where $a^{(0)}_0$ and $\tilde{a}^{(0)}_0$ are given by Eqs. \ref{a0}, and
\bea
a^{(0)}_1 &=& \int_0^1 \frac{dz}{{z}^{1-2 \Delta}} F_0^{(0)} \, \frac{(2\Delta-1) z^{2\Delta} - (2\Delta+1) z^{2\Delta-1} +2}{z^2+z+1}  \, {F}_0^{(0)} , \nonumber \\
\tilde{a}^{(0)}_1 &=& \int_0^1 \frac{dz}{{z}^{2-2 \Delta}} \tilde{F}_0^{(0)} \,\frac{(2\Delta-1) z^{2\Delta} - (2\Delta+1) z^{2\Delta-1} +2}{z^2+z+1}   \, {F}_0^{(0)}. 
\eea
Mulitplying the above by $\mu_c^2$, we get for the condensate
\be
\langle {\cal O}_{\Delta} \rangle^{(0)} = \gamma \,\,T^{\Delta} \sqrt{1-\frac{T^2}{T_c^2}} \left[ 1 + \mathcal{O} \left( \frac{1}{Q^{2\Delta -1}} \right) \right], \ee
where
\be \gamma =
\left(\frac{4\pi}{3}\right)^{\Delta} \sqrt{\Delta(4\Delta^2-1)}
\left(\frac{\frac{\Gamma^2\left(\tfrac{\Delta}{3}\right)}{\Gamma\left(\tfrac{2\Delta}{3}\right)} \, \frac{\Gamma\left(\tfrac{2(3-\Delta)}{3}\right)}{\Gamma^2\left(\tfrac{3-\Delta}{3}\right)} a^{(0)}_0 - \tilde{a}^{(0)}_0 }{\frac{\Gamma^2\left(\tfrac{\Delta}{3}\right)}{\Gamma\left(\tfrac{2\Delta}{3}\right)} \, \frac{\Gamma\left(\tfrac{2(3-\Delta)}{3}\right)}{\Gamma^2\left(\tfrac{3-\Delta}{3}\right)} a^{(0)}_1 - \tilde{a}^{(0)}_1} \right)^{1/2}~.
\ee

\begin{figure}[htbp]
\begin{center}
\includegraphics[width= 18 cm, keepaspectratio=true]{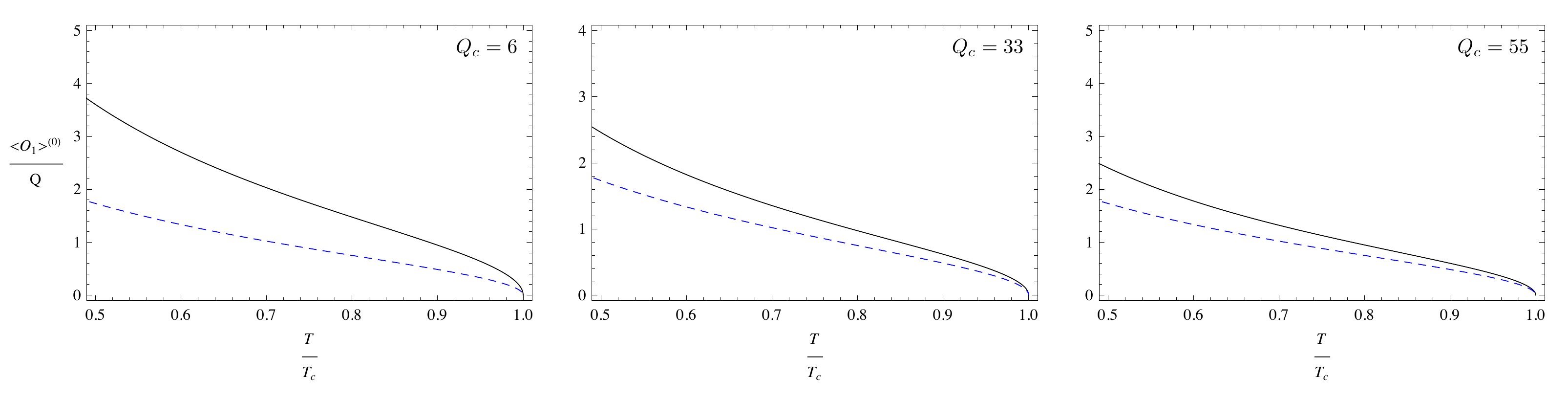}
\caption{The expectation value of the condensate for $\delta =1$ and $\Delta=1$ for $Q_c = 6$, $33$ and $55$. The solid black lines and the dashed blue lines depict the numerical and analytical results, respectively.}
\label{compare3}
\end{center}
\end{figure}

The expectation value of the condensate for $\delta =1$ and $\Delta =1$ obtained analytically is compared to the numerical results in Fig. \ref{compare3}. 

We can also obtain the higher modes of the condensate using a similar calculation as the one in Section \ref{highmode}, but with the electrostatic potential given in Eqs. \ref{leadinggaugeneartc} and \ref{gaugehighneartc}. The results are
\be
\frac{\langle {\cal O}_{\Delta}\rangle^{(1)}}{\langle {\cal O}_{\Delta}\rangle^{(0)}} = \frac{\delta \,(1-\delta)}{3-2\Delta} \left[1- \frac{\Gamma\left(\tfrac{5-2\Delta}{2}\right) \, \Gamma\left(\tfrac{3-2\Delta}{2}\right)\,\Gamma\left(2\Delta-1\right)}{\pi \, 2^{2\Delta -3}}\, \sin\left(\tfrac{\pi (3-2\Delta)}{2}\right) \right]\frac{\mu^2}{Q^2} \left[ 1 + {\cal O}\left(\frac{T}{Q}\right) \right] , 
\ee
and 
\be
\frac{\langle {\cal O}_{\Delta}\rangle^{(2)}}{\langle {\cal O}_{\Delta}\rangle^{(0)}} = \frac{\delta^2}{4 (3-2\Delta)} \left[\frac{\Gamma\left(\tfrac{5-2\Delta}{2}\right) \, \Gamma\left(\tfrac{3-2\Delta}{2}\right)\,\Gamma\left(2\Delta-1\right)}{\pi \, 2^{2\Delta -1}}\, \sin\left(\tfrac{\pi (3-2\Delta)}{2}\right) - \frac{1}{4}\right]\frac{\mu^2}{Q^2} \left[ 1+ {\cal O}\left(\frac{T}{Q}\right) \right]~. 
\ee
As before, for $\Delta = 1$, the ${\cal O} (\mu^2/Q^2)$ terms vanish, and the $n = 1$, $2$ modes are suppressed by a factor of an order ${\cal O} (\mu^2 \,T/Q^3)$ compared to the $n = 0$ mode. For $\delta = 1$, the odd modes for the condensate vanish.

\subsection{$T \ll T_c$}
Lastly, let us consider a low temperature  where the right hand side of Eq.\ (\ref{backreact}) is a lot larger than unity. We then have
 \be
A^{(1)} = \frac{{\langle {\cal O}_{\Delta} \rangle^{(0)}}^{\frac{1}{2\Delta}}\,2^{\frac{2\Delta-1}{2\Delta}}}{\Gamma\left(\tfrac{1}{2\Delta}\right) \, r_+^{\frac{1}{2}} \, \Delta^{\frac{1}{2\Delta}}} \, \sqrt{z} \,\, K_{\frac{1}{2\Delta}}\left(\frac{\langle {\cal O}_{\Delta} \rangle^{(0)} \, z^{\Delta}}{\Delta \, r_+^{\Delta}} \right) = A^{(0)} .
\ee
Substituting this into the equation of motion for the scalar and demanding regularity at the horizon, we again obtain Eq.\ (\ref{rovermudef}), but with
\bea
{\cal A} &=& \left(1-2\delta + \frac{3 \delta^2}{2} \right) {A^{(0)}}^2 \nonumber \\
&=& \left(1-2\delta + \frac{3 \delta^2}{2} \right) \,\frac{{\langle {\cal O}_{\Delta} \rangle^{(0)}}^{\frac{1}{\Delta}}\,2^{\frac{2\Delta-1}{\Delta}}}{\Gamma^2\left(\tfrac{1}{2\Delta}\right) \, \Delta^{\frac{1}{\Delta}}} \,\frac{z}{r_+}\, K^2_{\frac{1}{2\Delta}}\left(\frac{\langle {\cal O}_{\Delta} \rangle^{(0)} \, z^{\Delta}}{\Delta\,r_+^{\Delta}} \right).
\eea
Furthermore, since $\frac{1}{3} < \frac{1}{2\Delta} < 1$ is not an integer, we can express the modified Bessel function in terms of exponential and hypergeometric functions. The integral $a(1)$ then becomes
\bea
a(1) &=& \left(1-2\delta + \frac{3 \delta^2}{2} \right) \int_0^1 \frac{dz}{z^{2-2\Delta}} \, \frac{1}{h} \,  e^{-\tfrac{2 \langle {\cal O}_{\Delta} \rangle^{(0)} z^{\Delta}}{\Delta \,r_+^{\Delta}}} \Bigg[F_0^{(0)} \, _1F_1\left(\frac{1}{2}-\frac{1}{2\Delta};1-\frac{1}{\Delta};\frac{2  \langle {\cal O}_{\Delta} \rangle^{(0)} z^{\Delta}}{\Delta \, r_+^{\Delta}}\right) \Bigg]^2 \nonumber \\
&\approx&\left(1-2\delta + \frac{3 \delta^2}{2} \right) \, \frac{ \Gamma \left(\tfrac{2\Delta-1}{\Delta}\right)}{2^{\tfrac{2\Delta-1}{\Delta}}\Delta^{\tfrac{1-\Delta}{\Delta}}} \, \left(\frac{r_+^{\Delta} }{\langle {\cal O}_{\Delta} \rangle^{(0)}}\right)^{\tfrac{2\Delta-1}{\Delta}},
\eea
where we have approximated the integral by evaluating the integrand at $z=0$. Similarly, we find $\tilde{a}(1)$ to be of order $\left(\langle {\cal O}_{\Delta} \rangle^{(0)}\right)^{-2/\Delta}$, which is negligible compared to $a(1)$. Therefore,
\be
\frac{r_+^2}{\mu^2} = \frac{1-2\delta + \tfrac{3 \delta^2}{2}}{3 - 2\Delta}\, \frac{\Gamma^2\left(\tfrac{\Delta}{3}\right)}{\Gamma\left(\tfrac{2\Delta}{3}\right)} \, \frac{\Gamma\left(\tfrac{2(3-\Delta)}{3}\right)}{\Gamma^2\left(\tfrac{3-\Delta}{3}\right)} \, \frac{ \Gamma \left(\tfrac{2\Delta-1}{\Delta}\right)}{2^{\tfrac{2\Delta-1}{\Delta}}\Delta^{\tfrac{1-\Delta}{\Delta}}} \, \left(\frac{r_+^{\Delta} }{\langle {\cal O}_{\Delta} \rangle^{(0)}}\right)^{\tfrac{2\Delta-1}{\Delta}}. \label{lowtcondgen}
\ee
Multiplying by $\mu_c^2$, for $\delta =1$, we obtain the condensate
\be
\left[ \langle {\cal O}_{\Delta} \rangle^{(0)} \right]^{1/\Delta} = \left( 2 \,  \Delta^{\tfrac{1}{2\Delta-1}} \right)^{1-1/\Delta}\, \left[\frac{\Gamma\left(\tfrac{2\Delta-1}{\Delta}\right)}{\Gamma(2\Delta -1)}\right]^{\tfrac{1}{2\Delta-1}}\, \left( \frac{T_c}{T} \right)^{\frac{3-2\Delta}{2\Delta-1}} Q~,
\ee
showing again that the gap, $\langle \mathcal{O}_\Delta \rangle^{1/\Delta}$, is proportional to $Q$.


For $\delta \ne 1$, similarly, we obtain the condensate at low temperature,
\be
\left[ \langle {\cal O}_{\Delta} \rangle^{(0)} \right]^{1/\Delta} = \gamma \,\,T_c \left( \frac{T_c}{T} \right)^{\frac{3-2\Delta}{2\Delta-1}} \left[ 1 + \mathcal{O} \left( \frac{1}{Q^{2\Delta -1}} \right) \right] ,\ee
where
\be \gamma = \frac{4\pi}{3} \frac{1}{(2 \, \Delta^{\frac{1-\Delta}{2\Delta -1}})^{1/\Delta}} \, \left[\frac{\left(1-2\delta + \frac{3 \delta^2}{2}\right)\, \frac{\Gamma^2\left(\tfrac{\Delta}{3}\right)\,\Gamma\left(\tfrac{2(3-\Delta)}{3}\right) \, \Gamma \left(\tfrac{2\Delta-1}{\Delta}\right)}{\Gamma\left(\tfrac{2\Delta}{3}\right)\, \Gamma^2\left(\tfrac{3-\Delta}{3}\right)}}
{\frac{\Gamma^2\left(\tfrac{\Delta}{3}\right)}{\Gamma\left(\tfrac{2\Delta}{3}\right)} \, \frac{\Gamma\left(\tfrac{2(3-\Delta)}{3}\right)}{\Gamma^2\left(\tfrac{3-\Delta}{3}\right)} a^{(0)}_0 - \tilde{a}^{(0)}_0 }  \right]^{\frac{1}{2\Delta-1}}~,
\ee
where $a^{(0)}_0$ and $\tilde{a}^{(0)}_0$ are again defined in Eqs.\ (\ref{a0}).

We do not have a comparative plot for low temperature regime due to the fact that the numerical solution to the non-linear field equations become increasingly cumbersome as the temperature approaches zero.

By requiring regularity of the higher modes of the scalar field at the horizon, we can obtain the information concerning the higher modes of the condensate, as was done in Section \ref{highmode}. The results are
\be 
\langle {\cal O}_{\Delta} \rangle^{(1)} =  \frac{\delta \,(1-\delta)}{3-2\Delta} \frac{\Delta^{\tfrac{2-\Delta}{2}}}{2^{\tfrac{2}{\Delta}}} \, \Gamma\left(\frac{2}{\Delta}\right) \mu^2 \, \left[\langle {\cal O}_{\Delta} \rangle^{(0)}\right]^{1-\tfrac{2}{\Delta}},
\ee
and
\be 
\langle {\cal O}_{\Delta} \rangle^{(2)} =  \frac{\delta^2}{4(3-2\Delta)} \frac{\Delta^{\tfrac{2-\Delta}{2}}}{2^{\tfrac{2}{\Delta}}} \, \Gamma\left(\frac{2}{\Delta}\right) \mu^2 \, \left[\langle {\cal O}_{\Delta} \rangle^{(0)}\right]^{1-\tfrac{2}{\Delta}}.
\ee
It is easy to see that
\be
\frac{\langle {\cal O}_{\Delta}\rangle^{(n)}}{\langle {\cal O}_{\Delta}\rangle^{(0)}} = {\cal O}\Bigg(\left[\langle {\cal O}_{\Delta}\rangle^{(0)}\right]^{- \tfrac{2[n/2]}{\Delta}} \Bigg),
\ee
where $\left[n\right]$ denotes the largest integer $\leq n$, but with the odd modes vanishing for $\delta = 1$.

\section{Conclusion}
\label{sec:5}

In this article, we have discussed the properties of a striped holographic superconductor in the probe limit at large modulation wavenumber $Q$. We calculated the critical temperature $T_c$ and the expectation value of the condensate $\langle\mathcal{O}_\Delta \rangle$ below $T_c$ analytically for arbitrary values of the scaling dimension $\Delta$. 

We found that in the absence of a homogeneous terms in the chemical potential, both $T_c$ and $\langle\mathcal{O}_\Delta \rangle$ have a power law behavior for large $Q$. In particular, the critical temperature behaves as
\be
T_c \propto Q^{-\frac{2\Delta-1}{3-2\Delta}},
\ee
while the power of the condensate is such that the gap 
\be
\langle\mathcal{O}_\Delta \rangle^{1/\Delta} \propto Q. 
\ee
We also found that the odd modes of the condensate vanish, while the higher even modes are suppressed
\be
\frac{\langle {\cal O}_{\Delta}\rangle^{(n)}}{\langle {\cal O}_{\Delta}\rangle^{(0)}} \leq {\cal O}\left(\frac{\mu^{2n}}{Q^{2n}}\right).
\ee

In the case in which a homogeneous term is included in the chemical potential, both $T_c$ and $\langle\mathcal{O}_\Delta \rangle$ approach constant values in the large $Q$ limit, but the subleading terms are powers of $Q$. These constant values are the corresponding values for the homogeneous superconductors with chemical potential $\mu \delta$. We also found that the higher modes of the condensate are suppressed
\be
\frac{\langle {\cal O}_{\Delta}\rangle^{(n)}}{\langle {\cal O}_{\Delta}\rangle^{(0)}} \leq {\cal O}\left(\frac{\mu^{2 \left[n/2\right]}}{Q^{2 \left[n/2\right]}}\right).
\ee

The behavior of the gap indicates that the underlying mechanism does not rely on a correlation length, unlike in the weakly coupled BCS superconductors. The role of the correlation length is worth exploring further by calculating transport coefficients and correlation functions in the large $Q$ limit. Work in this direction is in progress.

In the weak coupling regime, including fluctuations results in a dip at the small $Q$ regime in which critical temperature decreases as $Q$ decreases pass a certain critical value \cite{Martin:2005fk}. We expect to see similar behavior once backreaction is included in the holographic calculation. Work in this direction is also in progress.

Lastly, we would also like to note that in the absence of homogeneous term in the chemical potential, the zero mode of the condensate is non-vanishing. In other words, the $\delta =1$ case does not correspond to the so-called ``pure" pair density wave (PDW) state. It will be interesting to find a holographic relaization of this pure PDW state and study its properties.

\acknowledgments{We wish to thank Stefanos Papanikolaou for illuminating discussions. J. H. is supported by West Virginia University start-up funds. The work of S.G., G.S., and J.T. is supported in part by the Department of Energy under grant DE-FG05-91ER40627.}

\bibliographystyle{kp}
\bibliography{References}
\end{document}